11-15-2021

# Library and Information Science Scholarly Journals Publishing Simulation: A Study

Priyanka Sinha
*Punjab University, Chandigarh*, priyankasinha101099@gmail.com

Subaveerapandiyan A
*Regional Institute of Education, Mysore*, subaveerapandiyan@gmail.com

# Library and Information Science Scholarly Journals Publishing Simulation: A Study

**Priyanka Sinha**

Ph.D. Research Scholar

Punjab University, Chandigarh

Email: priyankasinha101099@gmail.com

**Subaveerapandiyan A**

Professional Assistant (Library)

Regional Institute of Education Mysore, India

Email: subaveerapandiyan@gmail.com

**Abstract**

The author's productivity is assessed based on publications, which requires a lot of motivation and time. Manuscripts get through several steps before being accepted and published. The purpose of this paper is to understand the time gap between acceptance to the publication of manuscripts in reputed journals of Library and Information Science.

This paper is useful to contemporary researchers for knowing the journal publication duration. In this paper, we discussed the refereed and index journals in the field of library and information science. For this study, we collected the data from six LIS journals which were published from the 2020 January to December Asian region. The study focuses on detailed analyses of journal processing and publishing duration. The major contribution of this study gives the six LIS journal processing time they are: author manuscript submitted to accepted, accepted to published, and submitted to published period.

## 1. Introduction

Peer review is the term used in research publications and it is used for the quality of the papers in research writings. The duties and responsibilities of the reviewers are checking the papers thoroughly and giving updates to the editors. Peer reviews are mostly well experienced in research in specific fields of study. They check every part of the article such as title, abstract, keywords, introduction, objectives, scope and limitations, the significance of the study, review of literature, method, samplings, population, tools, originality of the data and information, coherence in the sentence, grammar, spelling, analysis, discussion, the conclusion is based on evidence and references. Peer-reviewed articles are called refereed journals. These refereed journals are abstracted and indexed in various prominent databases and websites such as Scopus, Web of Science, MEDLINE, PubMed, ERIC, EBSCO, DOAJ, BIOSIS, etc. The peer-review process is a critical step in research publication to ensure the quality of the manuscript and prove its scientific merit. The first-hand review process goes through the editor, where they decide the relevance of a particular article in a context journal. Before the actual review, the editors have the right to decide the acceptance or rejection of that paper. Editors also decide the potential reviewer and after continuous evaluation, the paper gets ready for publication. So, the key components of any manuscript are the author, editor, and reviewer. The duration in the whole process depends on these three key components, to understand the process and optimize the duration this study is being performed.

### 1.1 About Scimago

Scimago is used to measure the journals' qualities and citations based on that provide the details of Journal Rankings, country ranking, and subject ranking. It filtered options Open Access journals, Web of Science indexed, and Scientific Electronic Library Online (SciELO). Furthermore, it gave the Q1, Q2, Q3, Q4 rankings for the journals. Top 25% ranking in the same field of journals comes Q1 ranking. All the Scimago journals are indexed in the Scopus database. Scimago gave the details of the H-index of the journals, comprehensive documents published in a specific year, the total document cited in three years, and journal country details.

**1.2 Checklist for before submitting the manuscripts?**

- ➢ Check the scope of the journal before submitting the manuscripts
- ➢ Check the journal is indexed and Cross verify it is indexed in mentioned databases
- ➢ Check the frequency of publication
- ➢ Check the access type, whether it is open access or closed access
- ➢ Check the word limitation, referencing and citing formats, font style
- ➢ Check they will publish print, online or both
- ➢ Check the journal audience
- ➢ Check the peer review process
- ➢ Check the impact factor of the journal
- ➢ Check their publication speed

**1.3 Major reasons for manuscript rejections**

- ❖ Lack of research methods and experiments
- ❖ Lack of interpretations
- ❖ Lack of novelty
- ❖ Lack of coherence in sentence/abstract/introduction/research question/conclusion
- ❖ Inappropriate study
- ❖ Incomplete submission/statistics/conclusion
- ❖ Obsolescence literature and research study
- ❖ Already existing study
- ❖ Did not follow the length of the articles
- ❖ Poor Writing Skills (Language, structure, spelling, and grammatical errors)
- ❖ Wrong formatting of the manuscripts (Font size, style, citations, references, length of article, abstract, keywords, etc)
- ❖ Out of the scope of the journal
- ❖ Weak arguments
- ❖ Plagiarism

❖ Fake data and artificial results

**2. Objectives of the Study**

➢ To know the Library and information science journal publications duration

➢ To know the Scopus indexed Asian Library and Information Science Journals

➢ To understand the peer review type

➢ To understand the publishing periodicity

**3. Review of Literature**

Bilalli et. al. (2021) tried to figure out the peer review process duration in computer science journals. The study revealed that there was a significant difference between actual response time by publishers and data publicly available. Based on their study it was clear that the peer-review process took much more time than expected and the journal response time for computer science took a long revision time. This all concluded from the study that there was a wide gap between average first revision times reported and report by journal response time.

Hanafizadeh & Shaikh (2021) tried to explore the workflow of the peer-review process, the significant pitfalls, and the differences between micro and macro peer review processes. The backbone of any scientific publication is the peer-review process, where peer-reviewers perform a major role in ensuring quality publication and supporting decision-making. They used a purposive sample with an open-ended questionnaire on 33 candidates. The findings of the study show the peer-review process has a crucial role in scientific acceptance and publication. Further examination reveals that journals follow various models for this process, i.e., single reviewed, double-blind, and open-review models. The study suggests that the duration of the peer-review process and publication varies from journal to journal.

Bahadoran et al. (2020) discussed factors to choose a journal. In their article, they gave some valuable points for selecting appropriate journals. The criteria are indexed journals in standard databases, peer review, and process, impact factor, reputation published and editorial board, adopting publication ethics, publication periodicity, timeline and quality, article publication and processing charges, open access or subscription journals, print vs online journals or both.

Spezi et. al. (2018) tried to examine the peer review practice by open-access mega-journals, for this purpose they interviewed 31 senior publishers and editors from 16 different organizations. The purpose of the study was to highlight the weakness in the evaluation process and explore differences between realistic assessment and practical approaches. Findings show that reviewers represent the referee report in the same manner as conventional journals.

Huisman & Smits (2017) evaluated 3500 review experiences of authors to determine the quality and duration of the peer-review process. They used the website SciRev.sc to analyze the author's experiences of the review process, the time lag between the first review round and total review duration, and the editor's time to inform the author about acceptance and rejection. The first response time significantly varies from discipline. Total review duration is also needed to be considered other than first response time as these are different. Data reveals that there is a negative correlation between total review duration and journal impact factor. Immediate rejection time by editors is the reason behind unwanted time loss in the publication process. The study clearly shows that editors are not solely responsible for delays in publication, but manuscript handling at the editor's offices is equivalently accountable for the delay. The study also indicates that referee reports are difficult to understand of high-impact journals, as concluded highly ranked journals review process causes more damage to authors productivity.

Cooke et al. (2016) analyzed the peer review process and suggested reducing the duration of publication, for this purpose they have surveyed 6547 respondents published between 2012 to 2013 in biodiversity and conservation science disciplines. The key competency of any scientific publication is timeliness, this would extend the benefit of discoveries without obsolescence of data. The study revealed that most of the respondents agreed that the slow review process will affect the area of policy and management and when the review duration is too long data become outdated. A field like natural sciences and biodiversity suffer a lot due to delays in the publication of major studies because endangered species at the rate of obsolescence require special attention. Most of the respondents agreed that the rate of the review process needs to be altered in the fields of higher scientific significance.

Nguyen et. al. (2015) tried to recollect the opinion of authors regarding the peer-review process. The purpose of this survey was to analyze the consequences faced by authors and recipient communities due to delays in the peer-review process. For this survey authors collected responses from 6547 respondents via an online questionnaire. The respondents opined to have a short duration as compared to the usual process being followed. Most of the respondents felt that there was no long or short review period and the manuscript will be accepted or rejected. The respondent agreed that reviewers were responsible for review duration. The study also showed that respondents had suffered due to delays in the peer review process, lack of motivation, reduction of productivity, and potentially low-quality publication. The major outcome of the survey suggests that the review process can be altered by fixing deadlines, improving the journal management process, and better public outreach efforts.

Sabaj et. al. (2015) tried to explore the correlation between the duration of the peer review process, publication decision, and extent of agreement between reviewers. They estimated 369 peer review processes of three international Chilean journals, which were published in disciplines like humanities, engineering, and university teaching. The study shows that the peer review process took much more time than publication and there was a negative correlation between reviewer agreement and publication time. The results are helpful to understand the author's point of view, as the improvement is still needed in the peer review process and editors' contribution in reducing the duration of publication.

Björk & Solomon (2013) analyzed 135 peer-reviewed journals in various fields such as arts and humanities; business and economics; science technology and medical fields. They found that most science and technology field journals are reviewed faster than other fields. Furthermore, chemistry journals are published nine months, business and economic journals are published eighteen months. Additionally, they found electronic journals are published more fastly and paper submission delay, author revising the paper. These are also other reasons that publications delay not only a reviewer. Manuscript publications delay it does not only depend on the reviewer and editor includes author because of various editorial reasons.

### 4. Method

For this study we used observational method and secondary data was collected from six Library and Information Asian journals. They are DESIDOC Journal of Library and Information Technology (DJLIT), Annals of Library and Information Studies (ALIS), Journal of Information Science Theory and Practice (JISTAP), Journal of Information Science and Engineering (JLIS), Journal of Scientometric Research (JSCIRES) and Journal of Educational Media and Library Sciences (JOEMLS). We used Scimago website to identify these Asian Library Science journals. The search technique used we filtered the first subject area Social Sciences, second subject categories were chosen Library and Information Science, third chosen Asiatic Region and finally, we decided the year 2020. Twelve Asian Library and Information Science journals are listed in Scimago country rank, but we selected only six journals because the above mentioned journals only published the paper with timestamp. Further data was extracted from the official websites of each journal.

### 5. Data Analysis

**Table 1. Asian Library and Information Science Journals indexed in Scopus**

| S.No | Name of the Journal Based on SJR Ranking | Country | Indexed | APC |
|---|---|---|---|---|
| 1 | DESIDOC Journal of Library and Information Technology | India | Scopus/WoS | No |
| 2 | Malaysian Journal of Library and Information Science | Malaysia | Scopus/WoS | No |
| 3 | Annals of Library and Information Studies | India | Scopus/WoS | No |
| 4 | Journal of Information Science Theory and Practice | South Korea | Scopus | No |
| 5 | Journal of Library and Information Studies | Taiwan | Scopus/WoS | No |
| 6 | Journal of Information Science and Engineering | Taiwan | Scopus/WoS | Yes |
| 7 | Journal of Scientometric Research | India | Scopus/WoS | No |
| 8 | Journal of Educational Media and Library Sciences | Taiwan | Scopus | No |

| 9 | Pakistan Journal of Information Management and Libraries | Pakistan | Scopus | Yes |
|---|---|---|---|---|
| 10 | Wacana | Indonesia | Scopus/WoS | No |
| 11 | Library and Information Science | Japan | Scopus | No |
| 12 | IAFOR Journal of Literature and Librarianship | Japan | Scopus/WoS | No |

WoS-Web of Science, APC-Article Processing Charges

The above table 1 shows the Asian LIS journals, which were available in Scopus and Scimago. The total number of journals listed out of twelve in these ten journals are not collecting article processing charges, and eight journals are indexed in both Scopus and Web of Science. The journal origin of the countries: three journals from India and Taiwan; two journals from Japan and all other countries have one journal: Malaysia, South Korea, Pakistan, and Indonesia.

**Table 2. Timestamp Available Journals**

| S. No | Name of the Journal | Published Articles | Year | Periodicity | Review Type |
|---|---|---|---|---|---|
| 1 | DESIDOC Journal of Library and Information Technology (DJLIT) | 51 | 2020 | Bi-Monthly | Double-Blind |
| 2 | Annals of Library and Information Studies (ALIS) | 27 | 2020 | Quarterly | Double-Blind |
| 3 | Journal of Information Science Theory and Practice (JISTAP) | 24 | 2020 | Quarterly | Single-Blind |
| 4 | Journal of Information Science and Engineering (JLIS) | 84 | 2020 | Bi-Monthly | Double-Blind |
| 5 | Journal of Scientometric Research (JSCIRES) | 43 | 2020 | Triannual | Double-Blind |
| 6 | Journal of Educational Media and Library Sciences (JOEMLS) | 12 | 2020 | Triannual | Open Peer Review |

Table 2 shows six selective journals which have the timestamp in their published articles. These four journals are double-blind reviews, one journal is single-blind, and one more is open peer review. The periodicity of the journal is two journals are bi-monthly, and these two journals are published more than all other journals; equally, two journals are quarterly and triannual.

**Table 3. Submitted to Accepted Months/Year**

| S.No | Name of the Journal | <Month | <3 Months | <6 Months | <9 Months | <1 Year | >1 Year |
|---|---|---|---|---|---|---|---|
| 1 | DESIDOC Journal of Library and Information Technology (DJLIT) | 11.8 | 23.5 | 31.4 | 27.4 | 5.9 | - |
| 2 | Annals of Library and Information Studies (ALIS) | 6.2 | 6.2 | 18.8 | 56.3 | 12.5 | - |
| 3 | Journal of Information Science Theory and Practice (JISTAP) | - | 29.1 | 37.5 | 12.5 | 8.4 | 12.5 |
| 4 | Journal of Information Science and Engineering (JLIS) | 23.8 | 26.2 | 33.3 | 6 | 6 | 4.7 |
| 5 | Journal of Scientometric Research (JSCIRES) | - | 30.2 | 39.5 | 9.3 | 4.7 | 16.3 |
| 6 | Journal of Educational Media and Library Sciences (JOEMLS) | - | 8.3 | 58.4 | 25 | - | 8.3 |

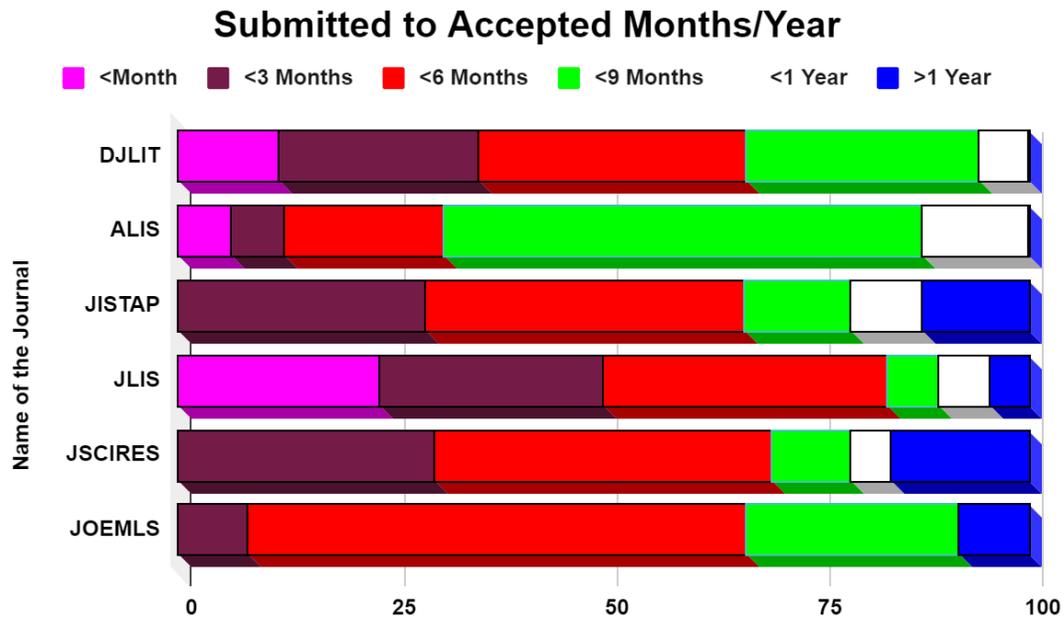

**Figure 1. Submitted to Accepted Months/Year**

Table 3 and Figure 1 portray articles submitted to accepted months and years. 50% of articles accepted less than six months are in these journals they are: DESIDOC Journal of Library and Information Technology, Journal of Information Science Theory and Practice took l, Journal of Information Science and Engineering, Journal of Scientometric Research, Journal of Educational Media and Library Sciences and more than six months took Annals of Library and Information Studies. In these journals of Educational Media and Library Sciences takes less duration for acceptance.

**Table 4. Accepted to Published Months/Year**

| S.No | Name of the Journal | <Month | <3 Months | <6 Months | <9 Months | <1 Year | >1 Year |
|---|---|---|---|---|---|---|---|
| 1 | DESIDOC Journal of Library and Information Technology (DJLIT) | 47 | 41.2 | 11.8 | - | - | - |
| 2 | Annals of Library and Information Studies (ALIS) | 50 | 43.8 | - | - | - | 6.2 |

| 3 | Journal of Information Science Theory and Practice (JISTAP) | 12.5 | 62.5 | 20.8 | - | - | 4.2 |
| 4 | Journal of Information Science and Engineering (JLIS) | - | - | 21.4 | 23.8 | 26.2 | 28.6 |
| 5 | Journal of Scientometric Research (JSCIRES) | 25.6 | 51.2 | 11.6 | - | 7 | 4.6 |
| 6 | Journal of Educational Media and Library Sciences (JOEMLS) | 41.7 | 50 | 8.3 | - | - | - |

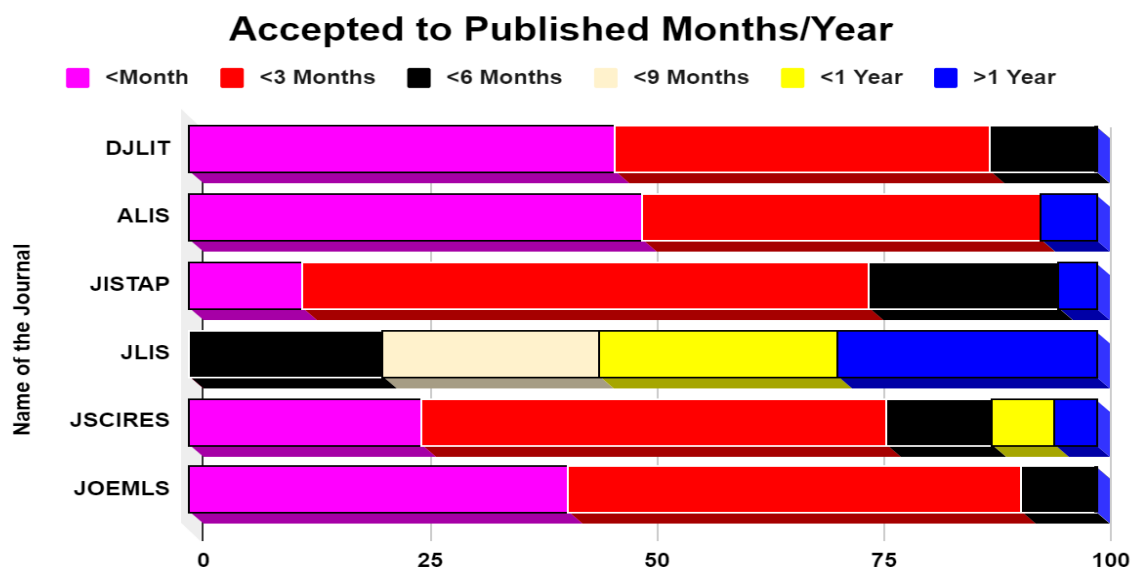

**Figure 2. Accepted to Published Months/Year**

The above table 4 and figure 2 reveals that the manuscript was accepted to be published months and years. Out of six journals, Annals of Library and Information Science journal published less than a month more than 50% of manuscripts and less than three months 43.8% manuscripts published. The Journal of Information Science and Engineering took more than three months to publish.

**Table 5. Submitted to Published Months/Year**

| S.No | Name of the Journal | <Month | <3 Months | <6 Months | <9 Months | <1 Year | >1 Year |
|---|---|---|---|---|---|---|---|
| 1 | DESIDOC Journal of Library and Information Technology | - | 17.6 | 37.3 | 35.3 | 7.8 | 2 |

| 2 | Annals of Library and Information Studies (ALIS) | - | 6.3 | 6.2 | 62.5 | 18.8 | 6.2 |
| 3 | Journal of Information Science Theory and Practice (JISTAP) | - | 8.4 | 20.8 | 33.3 | 16.7 | 20.8 |
| 4 | Journal of Information Science and Engineering (JLIS) | - | - | 17.8 | 1.2 | 26.2 | 54.8 |
| 5 | Journal of Scientometric Research (JSCIRES) | - | 14 | 27.9 | 32.6 | 4.6 | 20.9 |
| 6 | Journal of Educational Media and Library Sciences (JOEMLS) | - | - | 50 | 25 | 16.7 | 8.3 |

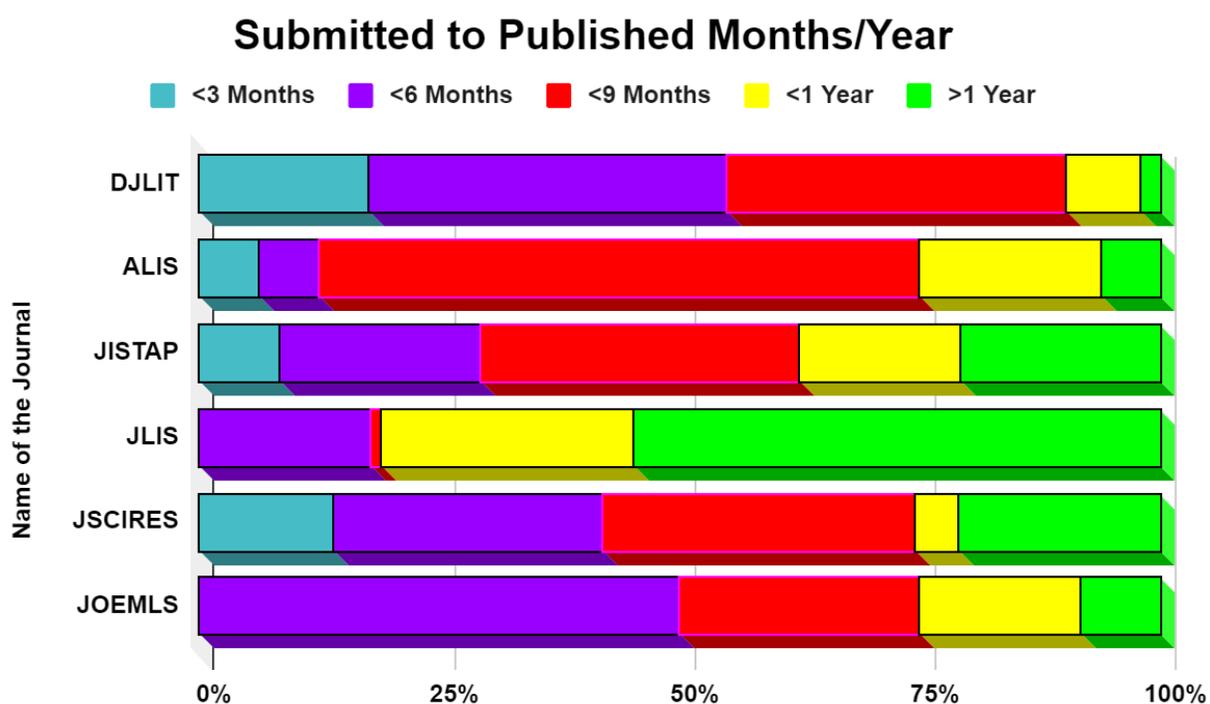

**Figure 3. Submitted to Published Months/Year**

Table 5 and figure 3 show that the manuscript was submitted to published months/year. DESIDOC Published less than six months more than 50% submitted journals and remaining all the journals take more than six months to publish.

**5. Conclusion**

Scholarly articles and research are increasing simultaneously but delay in peer-reviewed manuscript acceptance and publication has affected the productivity of research outcomes. Authors also have to wait for a long duration for the completion of the entire process, based on this study we can conclude that it takes a minimum of six months to nine months in LIS journals for the whole process and mostly follow the double-blind peer-review process. So, based on our study we can claim that most of the LIS journals take a longer duration from acceptance to publication. There was a three to six months gap between acceptance to publication as was evident from the data of DESIDOC Journal of Library and Information Technology. Hence, the value comparison suggests that the review process needs to be revised to reduce the overall delay and there should be some mechanism to optimize the outcome of the research. The literature study also suggests that reviewers should be held responsible for the review process and all these are affecting the productivity of authors. Literature also revealed that delay in the review process is not related to the certainty of acceptance or rejection. As per the study, we can conclude that to extend the benefit of research and scholarly communication, journals require transparency and improvisation in the whole process.